\documentclass{revtex4-2}
\usepackage[utf8]{inputenc}
\usepackage{amsbsy}
\usepackage{amsopn}
\usepackage{amstext}
\usepackage{amsmath,amsthm,amsfonts,amssymb}
\usepackage[mathcal]{eucal}
\usepackage{mathrsfs}
\usepackage{braket}
\usepackage[english]{babel}
\usepackage{color}
\usepackage{esint}
\usepackage{graphicx}
\usepackage{float}
\usepackage{units}
\usepackage{textcomp}
\usepackage{orcidlink}

\DeclareGraphicsExtensions{.png,PNG,.pdf,.PDF}
\usepackage{hyperref}
\usepackage{slashed}
\newcommand{\ie}{\begin{equation}}
\newcommand{\fe}{\end{equation}}
\newcommand{\se}{\begin{eqnarray}}
\newcommand{\ff}{\end{eqnarray}}
\begin{document}

\title{Dirac fermions in a spinning conical Gödel-type spacetime}
\author{R. R. S. Oliveira\,\orcidlink{0000-0002-6346-0720}}
\email{rubensrso@fisica.ufc.br}
\affiliation{Departamento de F\'isica, Universidade Federal do Cear\'a (UFC), Campus do Pici, C.P. 6030, Fortaleza, CE, 60455-760, Brazil}


\date{\today}

\begin{abstract}
In this paper, we determine the relativistic and nonrelativistic energy levels for Dirac fermions in a spinning conical G\"odel-type spacetime in $(2+1)$-dimensions, where we work with the curved Dirac equation in polar coordinates and we use the tetrads formalism. Solving a second-order differential equation for the two components of the Dirac spinor, we obtain a generalized Laguerre equation, and the relativistic energy levels of the fermion and antifermion, where such levels are quantized in terms of the radial and total magnetic quantum numbers $n$ and $m_j$, and explicitly depends on the spin parameter $s$ (describes the ``spin''), spinorial parameter $u$ (describes the two components of the spinor), curvature and rotation parameters $\alpha$ and $\beta$ (describes the conical curvature and the angular momentum of the spinning cosmic string), and on the vorticity parameter $\Omega$ (describes the G\"odel-type spacetime). In particular, the quantization is a direct result of the existence of $\Omega$ (i.e. such quantity acts as a kind of ``external field or potential''). We see that for $m_j>0$, the energy levels do not depend on $s$ and $u$; however, depend on $n$, $m_j$, $\alpha$, and $\beta$. In this case, $\alpha$ breaks the degeneracy of the energy levels and such levels can increase infinitely in the limit $\frac{4\Omega\beta}{\alpha}\to 1$. Already for $m_j<0$, we see that the energy levels depends on $s$, $u$ and $n$; however, it no longer depends on $m_j$, $\alpha$ and $\beta$. In this case, it is as if the fermion/antifermion ``lives only in a flat G\"odel-type spacetime''. Besides, we also study the low-energy or nonrelativistic limit of the system. In both cases (relativistic and nonrelativistic), we graphically analyze the behavior of energy levels as a function of $\Omega$, $\alpha$, and $\beta$ for three different values of $n$ (ground state and the first two excited states).
\end{abstract}

\maketitle

\section{Introduction}

In Relativistic Quantum Mechanics (RQM), the so-called Dirac fermions are spin-1/2 massive particles that obey the Dirac equation (DE), which is a relativistic wave equation derived by P. A. M. Dirac in 1928 \cite{Dirac1,Dirac2,Greiner,Bjorken,Grandy}. In this way, the spin-1/2 massive particles (or simply spin-1/2 particles) are also often so-called Dirac particles (e.g., electrons, protons, neutrons, quarks, muons, taus, and possibly neutrinos \cite{Studenikin,Lesgourgues}). So, in addition to the DE naturally explaining/proving the spin, helicity, and chirality of the spin-1/2 particles, predict that for each of these particles, there are their respective antiparticles (e.g., positrons, antiprotons, antineutrons,...) \cite{Greiner,Bjorken,Grandy,Martin,griffiths}. Already in Particle Physics, the Dirac fermions together with the bosons (integer spin particles) form the two fundamental classes of particles that make up the well-known Standard Model (SM) \cite{Martin,griffiths}. In particular, the DE (or Dirac fermions) has a wide variety of applications (somewhat incalculable), for example, it can be used to model (or study) the Dirac oscillator (a relativistic quantum harmonic oscillator for fermions) \cite{Moshinsky,Franco}, Aharonov-Bohm and Aharonov-Casher effects \cite{Aharonov,Hagen1,Hagen2,Oliveira1}, Aharonov-Bohm-Coulomb system \cite{Oliveira2,Oliveira3}, quantum rings \cite{Oliveira4}, ultracold and cold atoms \cite{Boada,Zhang}, trapped ions \cite{Lamata,Gerritsma1}, quantum Hall effect \cite{Schakel}, quantum computing \cite{Fillion,Huerta}, quantum phase transitions \cite{Bermudez1}, mesoscopic superposition states \cite{Bermudez2}, Zitterbewegung (trembling motion) \cite{Gerritsma1,Deriglazov}, Klein tunneling \cite{Casanova,Gerritsma2}, quark models \cite{Beveren,Becirevic}, Dirac materials (graphene, fullerenes, metamaterials, semimetals, and carbon nanotubes) \cite{Novoselov,Gonzalez,McCann,Ahrens,Armitage}, etc. Besides, the DE has already been worked with different types of potentials, such as the Coulomb potential \cite{Dong}, Yukawa potential \cite{Setare}, Eckart potential \cite{Zou}, linear potential \cite{Adame}, Woods–Saxon potential \cite{Guo}, P{\"o}schl-Teller potential \cite{Jia}, etc. Recently, the DE has been used to investigate the Dirac oscillator \cite{SousaOliveira,Oliveira5}, quantum Hall effect \cite{Oliveira6}, BTZ black hole \cite{Guvendi}, quantum interference and entanglement \cite{Ning}, electron-nucleus scattering \cite{Jakubassa}, Lorentz symmetry violation \cite{b1,b2}, rainbow gravity \cite{Ahmed,ARXIV}, etc.

In General Relativity (GR), one of the most relevant solutions is without a doubt the G\"odel metric (spacetime or G\"odel universe), found in 1949 by K. G\"odel \cite{Godel}, and which represents the first cosmological solution produced by rotating matter (or rotating universe). Some peculiar features of this solution are: is a regular/exact solution of the Einstein field equations (i.e. a singularity-free solution), has a negative cosmological constant and a non-null stress-energy tensor (sometimes called the stress–energy–momentum tensor), is stationary, homogeneous and anisotropic, has cylindrical symmetry, and consists of breaking the causality implying the possibility of closed time-like curves (CTCs), where such CTCs would allow time travels in a universe described by the G\"odel solution \cite{Deszcz,Gleiser,Barrow,Buser,Kerner}. In addition, the G\"odel original solution was generalized, among others, by M. J. Rebouças and J. Tiomno \cite{Rebouças}, who found all the spacetime homogeneous solutions of GR with rigid rotation, where such solutions are commonly called G\"odel-type solutions (or G\"odel-type universe), and are characterized by two real parameters: $l^2$ and $\Omega$ (both with the dimensions of length$^{-1}$), where $\Omega\geq 0$ is the vorticity (``rotation'') of the spacetime \cite{Figueiredo,Drukker,Boyda}. With respect to the values of $l^2$, such a parameter can assume three possible values: zero value ($l^2=0$), which corresponds to a flat solution (null curvature), a positive value ($l^2>0$), which corresponds to a hyperbolic solution (negative curvature), or a negative value ($l^2<0$), which corresponds to a spherical solution (positive curvature), respectively. However, for $l^2=\Omega^2/2$, we recover the G\"odel original solution (how should be) \cite{Godel}.

Besides, another solution of particular importance in GR is the metric (or spacetime) generated by a cosmic string, which is a kind of linear gravitational topological defect that can arise from a gauge theory with spontaneous symmetry breaking and perhaps may have been formed during the cosmological phase transitions in the early universe (Big Bang) \cite{Oli1,Oli2,Oli3,Oli4,INDIANA,Kibble,Vilenkin,Bezerra,Mello}. Some peculiar features of these cosmic defects are given as follows: are highly dense relativistic objects, stable, infinitely long and straight (unidimensional defects), can be static or spinning (with or without torsion), have cylindrical symmetry, are one of the exactly solvable models of GR, and the spacetime generated by them presents a locally flat geometry (but not globally) with a conical singularity (or non-trivial conical topology) at the origin characterized by a planar angular deficit, that is, a circle (or turn) around the cosmic string has an angle of less than 360° (therefore, they have a positive conical curvature or conical defects) \cite{Oli1,Oli2,Oli3,Oli4,INDIANA,Kibble,Vilenkin,Bezerra,Mello}. In that way, it says that cosmic strings have a conical curvature or singularity (or even a non-trivial conical topology) along their axis of symmetry ($z$-axis) \cite{Oli1,Oli2,Oli3,Oli4,INDIANA,Kibble,Vilenkin,Bezerra,Mello}. From an observational point of view, some detectors will try to look for signs of cosmic strings, for example, the Laser Interferometer Gravitational-Wave Observatory (LIGO) \cite{Cui}, the Laser Interferometer Space Antenna (LISA) \cite{Auclair}, and the North American Nanohertz Observatory for Gravitational Waves (NANOGrav) \cite{Blasi}. Now, from the point of view of condensed matter, there is a type of defect, so-called disclinations (in liquid crystals and crystalline solids \cite{Liu,Katanaev}) which have some interesting similarities with the cosmic strings, i.e. they are also linear topological defects with a conical singularity and break rotational symmetry of the medium \cite{Mello}. For example, for more information on conical graphene (or disclinations in graphene), or even graphene in the ``cosmic string spacetime'', we recommend Refs. \cite{Ardeshana,Vozmediano,Cortijo,Rozhkov}.

On the other hand, in recent years an interesting proposal has emerged to work with G\"odel-type solutions (or G\"odel-type metrics) in the presence of a cosmic string, which has resulted in various papers on the subject, mainly in the area of RQM \cite{Carv,Montigny,Garcia,Vitória,Eshghi,Ahmed1,Ahmed2,Sedaghatnia,GRG2024,Mustafa1,Mustafa2}. Explicitly, the general G\"odel-type metric (actually the line element) in the presence of a (static) cosmic string in cylindrical coordinates $(t,r,\phi,z)$ with signature $(+,-,-,-)$ is written as follows ($c=G=1$) \cite{Carv,Montigny,Garcia,Vitória,Eshghi,Ahmed1,Ahmed2,Sedaghatnia,GRG2024,Mustafa1,Mustafa2}
\begin{equation}\label{metric}
ds^2=\left(dt+\alpha\Omega\frac{(\sinh lr)^2}{l^2}d\phi\right)^2-dr^2-\alpha^2\frac{(\sinh 2lr)^2}{4l^2}d\phi^2-dz^2,
\end{equation}
where $0\leq r<\infty$ is the polar radial coordinate (or polar radius), with $r=\sqrt{x^2+y^2}$, $0\leq \phi\leq 2\pi$ is the polar angular coordinate (or polar angle), $-\infty<(t,z)<\infty$ are the temporal and axial coordinates, $\Omega$ is the vorticity parameter (``rotation parameter''), and $\alpha=1-4M$ ($0<\alpha\leq 1$) is called (conical) curvature or topological parameter (``angular deficit or deficit angle'') of the cosmic string with a linear mass density (mass per unit length) given by $M\geq 0$, respectively. In particular, for $\Omega=0$ (null vorticity) we obtain the ordinary or usual cosmic string spacetime ($\alpha\neq 1$) or the (flat) Minkowski spacetime ($\alpha=1$). Besides, it is well known that in the asymptotic limit $l\to 0$ (with $\alpha\to 1$) the metric above has the same geometry as the Som-Raychaudhuri metric (modeled only by $\Omega$), that is, the flat G\"odel-type spacetime is also often called the Som-Raychaudhuri spacetime, in which it was obtained by M. M. Som and A. K. Raychaudhuri in 1968 (therefore, the limit $l\to 0$ is also known as the Som-Raychaudhuri limit) \cite{Som}. It is also interesting to mention that this flat metric has appeared several times in string theory (not to be confused with cosmic strings) \cite{Drukker,Horowitz,Russo1,Russo2}, where have been reinterpreted as a G\"odel-type solution in string theory \cite{Boyda,Harmark}.

However, all Refs. \cite{Carv,Montigny,Garcia,Vitória,Eshghi,Ahmed1,Ahmed2,Sedaghatnia,GRG2024,Mustafa1,Mustafa2} have one thing in common (a ``limitation''), that is, they worked with the G\"odel-type spacetime only with a spinless/static cosmic string, that is, with a cosmic string without rotation or intrinsic angular momentum (spin). Thus, to ``close this gap'', here, we take into account the angular momentum of the cosmic string (i.e. here we consider a rotating or spinning cosmic string). Furthermore, another motivation for including the angular momentum of the cosmic string is that such a parameter is equivalent to the term inserted in the metric to study gravitational anyons via $(2+1)$-dimensional Einstein's gravity \cite{Bezerra,Cho}. So, according to Refs. \cite{Bezerra,Oli4}, to include the angular momentum of the cosmic string it is necessary to modify the temporal coordinate as $dt\to dt+\beta d\phi$, where $\beta\equiv 4J$ (with $c=G=1$) is a rotation parameter, and $J\geq 0$ is the linear density of angular momentum (i.e. momentum angular/spin per unit length along its axis of symmetry, which is the $z$-axis). Therefore, with this, we have from \eqref{metric} the following line element for the G\"odel-type spacetime in the presence of a spinning cosmic string
\begin{equation}\label{spinningmetric}
ds^2=\left(dt+\left(\beta+\alpha\Omega\frac{(\sinh lr)^2}{l^2}\right)d\phi\right)^2-dr^2-\alpha^2\frac{(\sinh 2lr)^2}{4l^2}d\phi^2-dz^2,
\end{equation}

In the present paper, we determine the relativistic and nonrelativistic energy levels for Dirac fermions in a spinning conical G\"odel-type spacetime (flat G\"odel-type spacetime with a spinning cosmic string) in $(2+1)$-dimensions ($dz^2=0$). However, as we now have two rotation parameters for the system, here we will always call $\Omega$ the vorticity parameter and $\beta$ simply the rotation parameter (this way we will avoid confusion). By way of illustration (a simplistic analogy), we can see our problem (general background) as a ``spinning top-tornado system'' (i.e. a ``spinning top living inside a tornado''), where the spinning top would be the spinning cosmic string and the tornado would be the Gödel-type spacetime (this analogy even makes ``a lot of sense'' because we can also consider the conical geometry of the spinning top as being the conical curvature of the cosmic string). So, to achieve our goal, we work with the curved DE (DE in a curved spacetime) in polar coordinates $(t,r,\phi)$, where the formalism used to obtain the solutions of this equation was the tetrads formalism of GR, which is written in terms of called spin connection. We chose this formalism because it is considered an excellent tool for studying Dirac fermions in curved spacetimes. In addition, here we also consider the ``spin'' of the (2D) planar fermion, described by a parameter $s=\pm 1$, called the spin parameter, where $s=+1(\uparrow)$ is for the ``spin up'', and $s=-1(\downarrow)$ is for the ``spin down'', respectively. In particular, this parameter arose from the study of the scattering of relativistic Dirac particles in an Aharonov-Bohm potential \cite{Hagen1}, and as a result of an exact equivalence between the Aharonov-Bohm effect and the Aharonov-Casher effect (both for Dirac particles) \cite{Hagen2}. Besides, for a more detailed analysis of an alternative approach to the dimensional reduction of the DE, we recommend Ref. \cite{Angelone}, where the authors performed a reduction from three to two spatial dimensions of the physics of a spin-1/2 fermion coupled to the electromagnetic field by applying Hadamard’s method of descent.

The structure of this paper is organized as follows. In Sect. \ref{sec2}, we introduce the curved DE in a spinning conical G\"odel-type spacetime, and using the tetrads formalism, we obtain a second-order differential equation for the two components of the Dirac spinor. In Sect. \ref{sec3}, we solve exactly and analytically this differential equation via a change of variable and the asymptotic behavior. Consequently, we obtain with this a generalized Laguerre equation, and also a quadratic polynomial equation for the total relativistic energy. Solving this polynomial equation, we obtain the relativistic energy levels (or high-energy spectrum). In Sect. \ref{sec4}, we analyze the nonrelativistic limit (low-energy limit), where we obtain the nonrelativistic energy levels (or low-energy spectrum). In both cases (relativistic and nonrelativistic), we graphically analyze the behavior of the spectrum as a function of the parameters $\Omega$, $\alpha$, and $\beta$ for three different values of $n$ (or three different quantum states). Finally, in Sect. \ref{sec5} we present our conclusions and some future perspectives. For simplicity, here we use the natural units $(\hslash=c=G=1)$, the spacetime with a signature $(+,-,-)$, and the Einstein notation (also known as the Einstein summation convention or Einstein summation notation).

\section{The curved Dirac equation in a spinning conical G\"odel-type spacetime \label{sec2}}

The covariant DE in a $(2+1)$-dimensional generic curved spacetime is given by the following equation (in polar coordinates) \cite{Oli1,Oli2,Oli3,Oli4,GRG2024,Greiner,Lawrie}
\begin{equation}\label{dirac1}
\left[i\gamma^\mu(x)\nabla_\mu(x)-m_0\right]\psi(t,r,\phi)=0, \ \ (\mu=t,r,\phi),
\end{equation}
where $\gamma^{\mu}(x)=e^\mu_{\ a}(x)\gamma^a$ are the curved gamma matrices, $\gamma^a=(\gamma^0,\gamma^1,\gamma^2)=(\gamma^0,\Vec{\gamma})$ are the flat gamma matrices, $e^\mu_{\ a}(x)$ are the tetrads, $\nabla_\mu(x)=\partial_\mu+\Gamma_\mu (x)$ is
the covariant derivative, $\partial_\mu=(\partial_t,\partial_r,\partial_\phi)$ are the partial derivatives, $\Gamma_\mu(x)=-\frac{i}{4}\omega_{ab\mu}(x)\sigma^{ab}$ is the spinorial connection (also called spinor affine connection), $\omega_{ab\mu}(x)$ is the spin connection, $\sigma^{ab}=\frac{i}{2}[\gamma^a,\gamma^b]=i\gamma^a\gamma^b$ ($a\neq b$) is a flat antisymmetric tensor, $m_0>0$ is the rest mass of the fermion, and $\psi=e^{\frac{i\phi\Sigma^3}{2}}\Psi_D$ is the two-component curvilinear spinor, where $\Psi_D\in\mathbb{C}^2$ is the original Dirac spinor. Here, the Latin indices $(a, b, c, \ldots)$ are used to label the local coordinates system (local reference frame or the Minkowski spacetime) and the Greek indices $(\mu, \nu, \alpha, \ldots)$ are used to label the general coordinates system (general reference frame or the curved spacetime), respectively.

Explicitly, Eq. \eqref{dirac1} can be rewritten as
\begin{equation}\label{dirac2}
\left\{i\gamma^t(x)\partial_t+i\gamma^r(x)\partial_r+i\gamma^\phi(x)\partial_\phi+i[\gamma^t(x)\Gamma_t(x)+\gamma^{r}(x)\Gamma_{r}(x)+\gamma^{\phi}(x)\Gamma_{\phi}(x)]-m_0\right\}\psi(t,r,\phi)=0.
\end{equation}

Now, we will focus our attention on the line element of the spinning conical G\"odel-type spacetime in $(2+1)$-dimensions as well as on the form of the metric, tetrads, curved gamma matrices, and spinorial and spin connections. Therefore, taking the limit $l\to 0$ (or $l^2=0$) in the line element \eqref{spinningmetric}, we have the following line element for the spinning conical G\"odel-type spacetime (i.e. making $\alpha\Omega r^2\to \beta+\alpha\Omega r^2$ in \cite{GRG2024} and, therefore, a generalization)
\begin{equation}\label{lineelement1}
ds^2=g_{\mu\nu}(x)dx^\mu dx^\nu=(dt+(\beta+\alpha\Omega r^2) d\phi)^2-dr^2-\alpha^2 r^2d\phi^2,
\end{equation}
where $g_{\mu\nu}(x)$ is the curved metric tensor (curved metric), whose inverse is given $g^{\mu\nu}(x)$, and are written as
\begin{equation}\label{metric1}
g_{\mu\nu}(x)=\left(\begin{array}{ccc}
1 & \ 0 & \beta+\alpha\Omega r^2 \\
0 & -1 &  0 \\
\beta+\alpha\Omega r^2 & \ 0 & (\beta+\alpha\Omega r^2)^2-\alpha^2 r^2
\end{array}\right), \ \
g^{\mu\nu}(x)=\left(\begin{array}{ccc}
\frac{\alpha^2 r^2-(\beta+\alpha\Omega r^2)^2}{\alpha^2 r^2} & 0 & \frac{(\beta+\alpha\Omega r^2)}{\alpha^2 r^2}\\
0 & -1 & 0\\
\frac{(\beta+\alpha\Omega r^2)}{\alpha^2 r^2} & 0 & -\frac{1}{\alpha^2 r^{2}}
\end{array}\right),
\end{equation}

So, with the line element well-defined (or metric well-defined), now we must build a local reference frame where the observer will be placed (the laboratory frame). In this way, we can define the gamma matrices in a curved spacetime \cite{Oli1,Oli2,Oli3,Oli4,GRG2024,Lawrie}. For example, using the tetrad formalism (of GR), it is perfectly possible to achieve this goal. In this formalism, a curved spacetime can be introduced point to point with a flat spacetime through objects of the type $e^\mu_{\ a}(x)$, which are called tetrads (square matrices), and which together with their inverses, given by $e^a_{\ \mu}(x)$, satisfy the following relations: $dx^\mu=e_{\ a}^\mu(x)\hat{\theta}^a$ and $\hat{\theta}^a=e^a_{\ \mu}(x)dx^\mu$, where $\hat{\theta}^a$ ($a=0,1,2$) is an object called noncoordinate basis. Furthermore, the tetrads and their inverses must also satisfy the following relations \cite{Oli1,Oli2,Oli3,Oli4,GRG2024,Lawrie}
\begin{eqnarray}\label{metric2}
&& g_{\mu\nu}(x)=e^a_{\ \mu}(x)e^b_{\ \nu}(x)\eta_{ab},
\nonumber\\
&& g^{\mu\nu}(x)=e^\mu_{\ a}(x)e^\nu_{\ b}(x)\eta^{ab},
\nonumber\\
&& g^{\mu\sigma}(x)g_{\nu\sigma}(x)=\delta^\mu_{\ \nu}=e^a_{\ \nu}(x)e^\mu_{\ a}(x),
\end{eqnarray}
where $\eta_{ab}=\eta^{ab}=$diag$(+1,-1,-1)$ is the Cartesian Minkowski metric tensor (flat metric or Lorentzian metric), which must satisfy
\begin{eqnarray}\label{metric3}
&& \eta_{ab}=e^\mu_{\ a}(x)e^\nu_{\ b}(x)g_{\mu\nu}(x),
\nonumber\\
&& \eta^{ab}=e^a_{\ \mu}(x)e^b_{\ \nu}(x)g^{\mu\nu}(x),
\nonumber\\
&& \eta^{ac}\eta_{cb}=\delta^a_{\ b}=e^a_{\ \mu}(x)e^\mu_{\ b}(x).
\end{eqnarray}

So, through the tetrad formalism, we can rewrite the line element \eqref{lineelement1} in terms of the noncoordinate basis as follows
\begin{equation}\label{lineelement2}
ds^2=\eta_{ab}\hat{\theta}^a\otimes\hat{\theta}^b=(\hat{\theta}^0)^2-(\hat{\theta}^1)^2-(\hat{\theta}^2)^2,
\end{equation}
where
\begin{equation}\label{bases}
\hat{\theta}^0=dt+(\beta+\alpha\Omega r^2) d\phi, \ \ \hat{\theta}^1=dr, \ \ \hat{\theta}^2=\alpha r d\phi, \ \ (dx^t=dt, \ dx^r=dr, \  dx^\phi=d\phi).
\end{equation}

In this way, the tetrads and their inverses are written as
\begin{equation}\label{tetrads}
e^{\mu}_{\ a}(x)=\left(
\begin{array}{ccc}
 1 & 0 & -\frac{(\beta+\alpha\Omega r^2)}{\alpha r} \\
 0 & 1 & 0 \\
 0 & 0 & \frac{1}{\alpha r} \\
\end{array}
\right), \ \
e^{a}_{\ \mu}(x)=\left(\begin{array}{ccc}
1 & 0 & (\beta+\alpha\Omega r^2) \\
0 & 1 & 0 \\
0 & 0 & \alpha r
\end{array}\right).
\end{equation}
where implies in the following curved gamma matrices
\begin{eqnarray}\label{gammamatrices}
&& \gamma^t(x)=\gamma^0-\frac{(\beta+\alpha\Omega r^2)}{\alpha r}\gamma^2,
\nonumber\\
&& \gamma^r(x)=\gamma^1,
\nonumber\\
&& \gamma^\phi(x)=\frac{1}{\alpha r}\gamma^2.
\end{eqnarray}

Now, we must find the form of the spin connection and later the spinorial connection. So, according to Ref. \cite{Lawrie}, this spin connection is defined as follows (torsion-free) 
\begin{equation}\label{spinconnection1}
\omega_{ab\mu}(x)=-\omega_{ba\mu}(x)=\eta_{ac}e^c_{\ \nu}(x)\left[e^\sigma_{\ b}(x)\Gamma^\nu_{\ \mu\sigma}(x)+\partial_\mu e^\nu_{\ b}(x)\right], 
\end{equation}
where $\Gamma^\nu_{\ \mu\sigma}(x)$ are the Christoffel symbols of the second type (symmetric tensor), and written as
\begin{equation}\label{Christoffel}
\Gamma^\nu_{\ \mu\sigma}(x)=\frac{1}{2}g^{\nu\lambda}(x)\left[\partial_\mu g_{\lambda\sigma}(x)+\partial_\sigma g_{\lambda\mu}(x)-\partial_\lambda g_{\mu\sigma}(x)\right]. 
\end{equation}

In that way, the non-null components for the Christoffel symbols are given by
\begin{eqnarray}\label{symbols}
&& \Gamma^{t}_{\ tr}=\Gamma^{t}_{\ rt}=\frac{(\beta\Omega+\alpha\Omega^2 r^2)}{\alpha r},\nonumber\\
&& \Gamma^{t}_{\ r\phi}=\Gamma^{t}_{\ \phi r}=\frac{(\beta+\alpha\Omega r^2)^2\Omega-\alpha\beta}{\alpha r},
\nonumber\\
&& \Gamma^{r}_{\ t \phi}=\Gamma^{r}_{\ \phi t}=\alpha\Omega r,
\nonumber\\
&& \Gamma^{r}_{\ \phi\phi}=(\beta+\alpha\Omega r^2)2\alpha\Omega r-\alpha^2 r,
\nonumber\\
&& \Gamma^{\phi}_{\ tr}=\Gamma^{\phi}_{\ rt}=-\frac{\Omega}{\alpha r},
\nonumber\\
&& \Gamma^{\phi}_{\ r\phi}=\Gamma^{\phi}_{\ \phi r}=\frac{(\alpha-\beta\Omega-\alpha\Omega^2 r^2)}{\alpha r}.
\end{eqnarray}

Consequently, the non-null components for the spin connection are written as
\begin{eqnarray}\label{spinconnection2}
&& \omega_{12t}=-\omega_{21t}=\Omega,
\nonumber\\
&& \omega_{02r}=-\omega_{20r}=\Omega,
\nonumber\\
&& \omega_{01\phi}=-\omega_{10\phi}=-\alpha\Omega r,
\nonumber\\
&& \omega_{12\phi}=-\omega_{21\phi}=\beta\Omega+\alpha\Omega^2 r^2-\alpha,
\end{eqnarray}
where implies in the following non-null components for the spinorial connection
\begin{eqnarray}\label{spinorialconnection}
&& \Gamma_t=-\frac{\Omega}{2}\gamma^1\gamma^2,
\nonumber\\
&& \Gamma_r=-\frac{\Omega}{2}\gamma^0\gamma^2,
\nonumber\\
&& \Gamma_\phi=\frac{(-\beta\Omega-\alpha\Omega^2 r^2+\alpha)\gamma^1\gamma^2+(\alpha\Omega r)\gamma^0\gamma^1}{2}.
\end{eqnarray}

So, using the expressions \eqref{gammamatrices} and \eqref{spinorialconnection}, we obtain the following contribution of the spinorial connection
\begin{equation}\label{contributionofthespinorialconnection}
[\gamma^t(x)\Gamma_t(x)+\gamma^{r}(x)\Gamma_{r}(x)+\gamma^{\phi}(x)\Gamma_{\phi}(x)]=\frac{\Omega}{2}\gamma^{0}\gamma^1\gamma^2+\frac{1}{2 r}\gamma^{1}.
\end{equation}

Therefore, using the expressions \eqref{contributionofthespinorialconnection} and \eqref{gammamatrices} in \eqref{dirac2}, we obtain the following DE a in spinning conical G\"odel-type spacetime
\begin{equation}\label{dirac3}
\left[i\gamma^0\partial_t+i\gamma^1\left(\partial_r+\frac{1}{2r}\right)+i\gamma^2\left(\frac{1}{\alpha r}\partial_\phi-\frac{(\beta+\alpha\Omega r^2)}{\alpha r}\partial_t\right)+\gamma^0\Vec{S}\cdot\Vec{\Omega}-m_0\right]\psi(t,r,\phi)=0,
\end{equation}
or in terms of the Dirac Hamiltonian $H_D$, such as
\begin{equation}
i\partial_t\psi(t,r,\phi)=H_D\psi(t,r,\phi), \ \ H_D=\left[-i\gamma^0\gamma^1\left(\partial_r+\frac{1}{2r}\right)-i\gamma^0\gamma^2\left(\frac{1}{\alpha r}\partial_\phi-\frac{(\beta+\alpha\Omega r^2)}{\alpha r}\partial_t\right)-\Vec{S}\cdot\Vec{\Omega}+\gamma^0 m_0\right],
\end{equation}
where the term $\Vec{S}\cdot\Vec{\Omega}=\frac{1}{2}\Omega\Sigma^3$ ($\Omega=\Omega_z=\Omega_3$) is a ``spin-vorticity coupling'', with $\Vec{S}=\frac{1}{2}\Vec{\Sigma}=(S^1,S^2,S^3)$ being the spin vector, $\Vec{\Omega}=\Omega\Vec{e}_z$ and $\Sigma^3=\Sigma_3=i\gamma^1\gamma^2=\Sigma^z$. In particular, this coupling is very similar to the spin-rotation coupling for Dirac fermions in rotating frames \cite{Oli1,Oli2,Oli3,Oli4}. Furthermore, as we will see later, the cause of the quantization of energy levels is the vorticity $\Omega$, such as the cause of the quantization of energy levels in rotating frames is the angular velocity $\omega$ or $\Vec{\omega}=\omega\Vec{e}_z$ (where the energy levels of both are even very similar). However, for both cases, the quantization does not arise because of these couplings, but rather due to a linear term in $r$ (our case is given by $-\Omega r \partial_t$, and for fermions in rotating frames, it is given by $\alpha\omega r \partial_t$ \cite{Oli1,Oli2,Oli3,Oli4}). Therefore, this implies a very strong link between vorticity and angular velocity (in fact, in fluid mechanics, vorticity is used to quantify the rotation of particles in a moving fluid).

In addition, assuming that our system is a stationary system (time-independent Dirac Hamiltonian), we can define an ansatz for the two-component spinor $\psi(t,r,\phi)$ as follows \cite{GRG2024,Oli4}
\begin{equation}\label{spinor}
\psi(t,r,\phi)=\frac{e^{i(m_j\phi-Et)}}{\sqrt{2\pi}}\left(
           \begin{array}{c}
            R_+(r) \\
            iR_-(r) \\
           \end{array}
         \right),
\end{equation}
where the components $R_+(r)$ and $R_-(r)$ are real radial functions (with $R_+(r)\neq R_-(r)$), $m_j=\pm 1/2,\pm 3/2, \pm 5/2,\ldots$ is the total magnetic quantum number, and $E$ is the relativistic total energy of the fermion (or of the system).

On the other hand, since we are working in $(2+1)$-dimensions, implies that the flat gamma matrices $\gamma^a=\eta^{ab}\gamma_b=(\gamma^0,\gamma^1,\gamma^2)=(\gamma_0,-\gamma_1,-\gamma_2)$ take the form: $\gamma_1=\sigma_3\sigma_1=i\sigma_2$,  $\gamma_2=s\sigma_3\sigma_2=-is\sigma_1$ ($s=\pm 1$ is the spin parameter), and $\gamma^0=\Sigma^3=\sigma_3$ \cite{Hagen1,Hagen2,GRG2024}, where the Pauli matrices are written as follows
\begin{equation}\label{Paulimatrices}
\sigma_1=\left(
    \begin{array}{cc}
      0\ &  1 \\
      1\ & 0 \\
    \end{array}
  \right), \ \  \sigma_2=\left(
    \begin{array}{cc}
      0 & -i  \\
      i & \ 0 \\
    \end{array}
  \right), \ \  \sigma_3=\left(
    \begin{array}{cc}
      1 & \ 0 \\
      0 & -1 \\
    \end{array}
  \right).
\end{equation}

In that way, we can obtain from \eqref{dirac3} the following equation 
\begin{equation}\label{DIRAC}
\left[\sigma_3 E+\sigma_2\left(\frac{d}{dr}+\frac{1}{2r}\right)-is\sigma_1\left(\frac{m_j+\beta E}{\alpha r}+\Omega E r\right)+\frac{\Omega}{2}-m_0\right]\left(
           \begin{array}{c}
            R_+(r) \\
            iR_-(r) \\
           \end{array}
         \right)=0,
\end{equation}
where the parameters $\alpha$ and $\beta$ have the goal of modifying (``shifting'') the total angular momentum of the fermion (along the $z$-axis) \cite{Bezerra}. Therefore, we can now define a ``new total magnetic quantum number'' (or an ``effective magnetic quantum number''), given by: $M_j=M^{eff}_j\equiv(m_j+\beta E)/\alpha$ ($\beta E=\vert \beta E\vert\geq 0$), where now the operator $J_z=J_z^{usual}=-i\partial_\phi$ ($z$-component of the total angular momentum operator $\Vec{J}$ in polar coordinates) is rewritten as $J_z^{eff}=-i(\partial_\phi-\beta\partial_t)/\alpha$ \cite{Bezerra,Cho}. So, comparing $J_z^{eff}$ with $J_z^{usual}$, we have an additional contribution for the total angular momentum of the fermion that takes account of the curvature and the angular momentum of the cosmic string \cite{Bezerra,Cho}.

So, from \eqref{DIRAC}, we have two coupled first-order differential equations for $R_+(r)$ and $R_-(r)$, given by
\begin{equation}\label{dirac4}
\left(m_0-\frac{\Omega}{2}-E\right)R_+(r)=\left[\frac{d}{dr}+sm_0\omega r+\frac{s}{\alpha r}\left(\bar{m}_j+\frac{s\alpha}{2}\right)\right]R_-(r),
\end{equation}
and
\begin{equation}\label{dirac5}
\left(m_0-\frac{\Omega}{2}+E\right)R_-(r)=\left[\frac{d}{dr}-sm_0\omega r-\frac{s}{\alpha r}\left(\bar{m}_j-\frac{s\alpha}{2}\right)\right]R_+(r),
\end{equation}
 where $\omega\equiv\frac{\Omega E}{m_0}=\frac{\vert\Omega E\vert}{m_0}\geq 0$ is a kind of ``angular frequency'' (or ``vorticity frequency''), and we define for convenience that $\Bar{m}_j\equiv m_j+\beta E$.

Therefore, substituting \eqref{dirac5} in \eqref{dirac4}, and later  \eqref{dirac4} in  \eqref{dirac5}, we obtain the following second-order differential equation for the DE in a spinning conical G\"odel-type spacetime
\begin{equation}\label{dirac6}
\left[\frac{d^2}{dr^2}+\frac{1}{r}\frac{d}{dr}-\frac{M^2_u}{\alpha^2 r^2}-(m_0\omega r)^2+E_u\right]R_u(r)=0,
\end{equation}
where we define
\begin{equation}\label{dirac7}
M_u\equiv\left(\Bar{m}_j-\frac{su\alpha}{2}\right), \ \ E_u\equiv E^2-\left(m_0-\frac{\Omega}{2}\right)^2-\frac{2m_0\omega}{\alpha}\left(\Bar{m}_j+\frac{su\alpha}{2}\right),
\end{equation}
where the parameter $u=\pm 1$ (``spinorial parameter'') represent the two components of the spinor: $u=+1$ is for the upper component and $u=-1$ is for the lower component, respectively.

\section{The relativistic energy levels\label{sec3}}

To solve exactly and analytically Eq. \eqref{dirac6}, first we will introduce a new (dimensionless) variable in the system, given by: $w=m_0\omega r^2\geq 0$. Thus, through a change of variable, Eq. \eqref{dirac6} becomes
\begin{equation}\label{dirac8}
\left[w\frac{d^{2}}{dw^{2}}+\frac{d}{dw}-\frac{M^{2}_u}{4\alpha^2w}-\frac{w}{4}+\frac{E_u}{4m_0\omega}\right]R_u(w)=0.
\end{equation}

Now, we need to analyze the asymptotic behavior of Eq. \eqref{dirac8} for $w\to 0$ and $w\to\infty$. Once done, we obtain a regular solution to this equation given as follows
\begin{equation}\label{dirac9}
R_u(w)=C_u w^{\frac{\vert M_u\vert}{2\alpha}}e^{-\frac{w}{2}}\bar{R}_u(w), \ \ (\vert M_u\vert>0),
\end{equation}
where $C_u>0$ are normalization constants, $\bar{R}_u(w)$ are unknown functions to be determined, and $R_u(w)$ must satisfy the following boundary conditions to be a normalizable solution (i.e. the solutions of the system must obey two boundary conditions to be physically consistent)
\begin{equation}\label{conditions} 
R_u(w\to 0)=0, \ \ R_u(w\to\infty)=0.
\end{equation}

So, substituting \eqref{dirac9} in \eqref{dirac8}, we have a second-order differential equation for $\bar{R}_u(w)$ given as follows
\begin{equation}\label{dirac10}
\left[w\frac{d^{2}}{dw^{2}}+(\vert\bar{M}_u\vert-w)\frac{d}{dw}-\Bar{E}_u\right]\bar{R}_u(w)=0,
\end{equation}
where
\begin{equation}\label{define}
\vert\bar{M}_u\vert\equiv\frac{\vert M_u\vert}{\alpha}+1, \ \ \Bar{E}_u\equiv\frac{\vert\bar{M}_u\vert}{2}-\frac{E_u}{4m_0\omega}.
\end{equation}

It is not difficult to note that Eq. \eqref{dirac10} is the well-known generalized Laguerre equation, whose solutions are the generalized Laguerre polynomials, written as $\bar{R}_u(w)=L^{\frac{\vert M_u\vert}{\alpha}}_n(w)$  \cite{Greiner}. Consequently, the quantity $\Bar{E}_u$ must be equal to a negative integer, i.e. $\Bar{E}_u=-n$ (a quantization condition), where $n=n_r=0,1,2,\ldots$ is a quantum number (also called radial quantum number). Therefore, from this condition, we obtain the following quadratic polynomial equation for the total relativistic energy $E$
\begin{equation}\label{dirac11}
E^2-4\Omega E\left[n+\frac{su}{2}+\frac{\vert \beta E+L_u\vert+(\beta E+L_u)}{2\alpha}\right]-\left(m_0-\frac{\Omega}{2}\right)^2=0,
\end{equation}
with $L_u=L_u(\alpha)\equiv m_j-\frac{su\alpha}{2}\gtrless 0$ being a ``topological quantum number'' (because it is a function on the topological parameter $\alpha$).

Therefore, solving the polynomial equation \eqref{dirac11} for $L_u>0$ and $L_u<0$, we obtain the following relativistic energy spectrum (relativistic energy levels) for Dirac fermions in a spinning conical G\"odel-type spacetime
\begin{equation}\label{spectrum}
E^\sigma_{n,m_j,s,u}=\frac{2\Omega N_u}{\left[1-\frac{4\Omega\beta}{\alpha}\left(\frac{\vert L_u \vert+L_u}{2L_u}\right)\right]}+\sigma\sqrt{\frac{4\Omega^2 N_u^2}{\left[1-\frac{4\Omega\beta}{\alpha}\left(\frac{\vert L_u \vert+L_u}{2L_u}\right)\right]^2}+\frac{\left(m_0-\frac{\Omega}{2}\right)^2}{\left[1-\frac{4\Omega\beta}{\alpha}\left(\frac{\vert L_u \vert+L_u}{2L_u}\right)\right]}},
\end{equation}
where we define
\begin{equation}\label{define}
N_u=N^{eff}_u(\alpha)=N^{total}_u(\alpha)=\left(n+\frac{1+su}{2}+\frac{\vert L_u\vert+L_u}{2\alpha}\right)\geq 0,
\end{equation}
being $\sigma=\pm 1$ a parameter (``energy parameter'') which represents the positive energy states/solutions (particle/fermion with energy $E_{particle}=E^+=\vert E^+\vert>0$) and the negative energy states/solutions (antiparticle/antifermion with energy $E_{antiparticle}=-E^-=\vert E^-\vert>0$ \cite{Greiner,Bjorken,Grandy}), and $N_u$ is a ``topological total or effective quantum number'' (since it depends on all other quantum numbers and also on $\alpha$). Additionally, a noteworthy observation is that the particle and antiparticle do not possess equal energies ($E_{particle}\neq E_{antiparticle}$), i.e. here the spectrum is not symmetrical. Therefore, this asymmetry (caused by extra terms outside the square root \cite{GRG2024}) in energy levels does not emphasize the equilibrium or equality between particle and antiparticle. So, we see that the spectrum \eqref{spectrum} explicitly depends on the quantum numbers $n$ and $m_j$, spin parameter $s$, spinorial parameter $u$, curvature or topological parameter $\alpha$, rotation parameter $\beta$, and on the vorticity parameter $\Omega$, respectively. In particular, the quantization (or discretization) of the spectrum is a direct result of the existence of $\Omega$, where such quantity acts as a kind of ``external field or potential''. Therefore, here we technically do not have a free fermion (whose spectrum in this case is continuous), but subject to a ``field or potential'' modeled by $\Omega$. In fact, in the absence of vorticity ($\Omega=0$), we obtain the spectrum of a truly free fermion/antifermion, whose spectrum is the rest energy $E_0^{\pm}=\pm m_0$ (i.e. it is the energy of a relativistic fermion/antifermion without any interaction). Thus, it implies that a spinning cosmic string by itself does not generate the quantization of the spectrum, that is, the spectrum is only quantized in terms of $n$ and $m_j$ because of $\Omega$.

So, comparing the spectrum \eqref{spectrum} (for $\beta=0$) with the spectrum of the literature for scalar bosons (Klein-Gordon or spin-0 particles), also in $(2+1)$-dimensions ($k=0$), that is, everything/both on an “equal footing”, we verify that our case (fermionic case) has some similarities and difference with the bosonic case. For example, according to Ref. \cite{Carv}, the spectrum for scalar bosons in the (flat) Gödel-type spacetime in the presence of a static cosmic string is given by (already including the spectrum of the antiparticle since this was ignored)
\begin{equation}\label{bosons}
E^\pm_{n,m}=E^{bosons}_{n,m}=2\Omega\left(n+\frac{1}{2}+\frac{\vert m\vert +m}{2\alpha} \right)\pm\sqrt{4\Omega^2\left(n+\frac{1}{2}+\frac{\vert m\vert +m}{2\alpha} \right)^2+M^2},
\end{equation}
where $m=0,\pm 1,\pm 2,\ldots$ is the orbital magnetic quantum number or simply the magnetic quantum number (is the same as what appears in the Schrödinger equation), and $M$ is the rest mass of the boson (or antiboson). In this way, we see that similar to the bosonic case, our case also depends on two quantum numbers, where the quantum number $n$ is the same for both cases. However, unlike the bosonic case, our second quantum number (given by $m_j$) has semi-integer and non-zero values; consequently, regardless of the value of $m_j$ (since $L_u\neq 0$), our spectrum will still depend on $\alpha$ (which does not happen for the bosonic case with $m=0$). Thus, similar to the bosonic case, our case also depends on a linear and quadratic term in $\Omega$; however, in our case, the mass $m_0$ is ``tied'' to $\Omega$ (in fact, this is a consequence of the spin-vorticity coupling, which does not happen for bosons). Therefore, this implies that for massless fermions ($m_0=0$), the spectrum of the antiparticle will still be non-zero, which does not happen for massless bosons ($M=0$), that is, $E^{antiparticle}_{n,m_j,s,u}(m_0=0)\neq 0$ (fermionic case) and $E^{antiparticle}_{n,m}(M=0)=0$ (bosonic case), respectively.

Besides, it is also interesting to analyze the spectrum according to the values $L_u$. Therefore, in Table \eqref{tab1} we have two possible settings for the spectrum depending on the values of $L_u$. So, according to Table \eqref{tab1}, we see that for $L_u>0$ (or $m_j>0$), where the angular momentum is positive (fermion orbiting parallel to the rotation of the cosmic string), the spectrum does not depend on $s$ and $u$, that is, the spectrum is exactly the same regardless of the spin or spinor component chosen ($\vert E_\uparrow^{\pm}\vert=\vert E_\downarrow^{\pm}\vert$); however, it depends on both $n$, $m_j$, $\alpha$ and also on $\beta$. Therefore, in this case, the parameters $\alpha$ and $\beta$ explicitly affect the values of the spectrum, and the ground state ($n=0$) still depends on $m_0$, $\Omega$, $\alpha$, and $\beta$ (i.e. both the curvature and rotation of the cosmic string, as well as the vorticity of G\"odel-type spacetime, affect all energetic states of the fermion). Furthermore, due to the presence of $\alpha$, the spectrum no longer has a well-defined degeneracy, in other words, the function of $\alpha$ (one of them) is to break the degeneracy of the spectrum \cite{Carv}. Indeed, in the absence of the cosmic string ($\alpha=1$ and $\beta=0$), then we have a spectrum with a well-defined degeneracy (finite or infinite, for example) \cite{Oli4}. On the other hand, it is also important to comment that the quantity $\frac{4\Omega\beta}{\alpha}$ (that depends on all parameters associated with the spinning conical Gödel-type spacetime) can make the spectrum diverge, that is, the spectrum can increase infinitely (actually have a very large amount of energy) in the limit $\frac{4\Omega\beta}{\alpha}\to 1$ (we will see this soon in graphs). In other words, depending on the vorticity values of the Gödel-type spacetime (modeled by $\Omega$), of the rotation (modeled by $\beta$) and curvature (modeled by $\alpha$) of the cosmic string, the energies can diverge. In particular, we can see this quantity as a consequence of a coupling between $\Omega$ and $\beta$, given by $\Vec{\Omega}\cdot\Vec{\beta}$ (with $\Vec{\beta}=\beta\Vec{e}_z$), in which we can call ``rotation-rotation coupling'', or ``vorticity-angular momentum coupling''. In that way, as $\Omega\beta\geq 0$, implies that $\Omega$ and $\beta$ are parallel, that is, they have the same direction of rotation (and is a counterclockwise or positive rotation). Already for $L_u<0$ (or $m_j<0$), where the angular momentum is negative (fermion orbiting antiparallel to the rotation of the cosmic string), the spectrum now depends on $s$ and $u$, as well as $n$, that is, the spectrum depends on the spin and spinor component chosen ($\vert E_{s,u}^{\pm}\vert=\vert E_{s,u}^{\pm}\vert$ for $su=+1$ and $\vert E_{s,u}^{\pm}\vert\neq\vert E_{s,u}^{\pm}\vert$ for $su=-1$); however, it no longer depends on both $m_j$, $\alpha$ and $\beta$. Therefore, in this case, the parameters $\alpha$ and $\beta$ no longer affect the values of the spectrum, and the ground state ($n=0$) depends only on $m_0$ and $\Omega$ (regardless of the value of the product $su$). Furthermore, due to the absence of $\alpha$ and $\beta$, it is as if the fermion/antifermion ``lives only in a flat G\"odel-type spacetime'', i.e. without the presence of a spinning cosmic string.

\begin{table}[h]
\centering
\begin{small}
\caption{Relativistic spectrum depends on the values of $L_u$.} \label{tab1}
\begin{tabular}{ccc}
\hline
Setting & $L_u$ & Spectrum \\
\hline
1& $L_u>0$ & \ \ $E^\sigma_{n,m_j}=\frac{2\Omega\left(n+\frac{1}{2}+\frac{m_j}{\alpha}\right)}{\left[1-\frac{4\Omega\beta}{\alpha}\right]}+\sigma\sqrt{\frac{4\Omega^2\left(n+\frac{1}{2}+\frac{m_j}{\alpha}\right)^2}{\left[1-\frac{4\Omega\beta}{\alpha}\right]^2}+\frac{\left(m_0-\frac{\Omega}{2}\right)^2}{\left[1-\frac{4\Omega\beta}{\alpha}\right]}}$\\
2& $L_u<0$ & \ \ $E^\sigma_{n,s,u}=2\Omega \left(n+\frac{1+su}{2}\right)+\sigma\sqrt{4\Omega^2\left(n+\frac{1+su}{2}\right)^2+\left(m_0-\frac{\Omega}{2}\right)^2}$\\
\hline
\end{tabular}
\end{small}
\end{table}

\begin{figure}[!h]
\centering
\includegraphics[width=10.0cm]{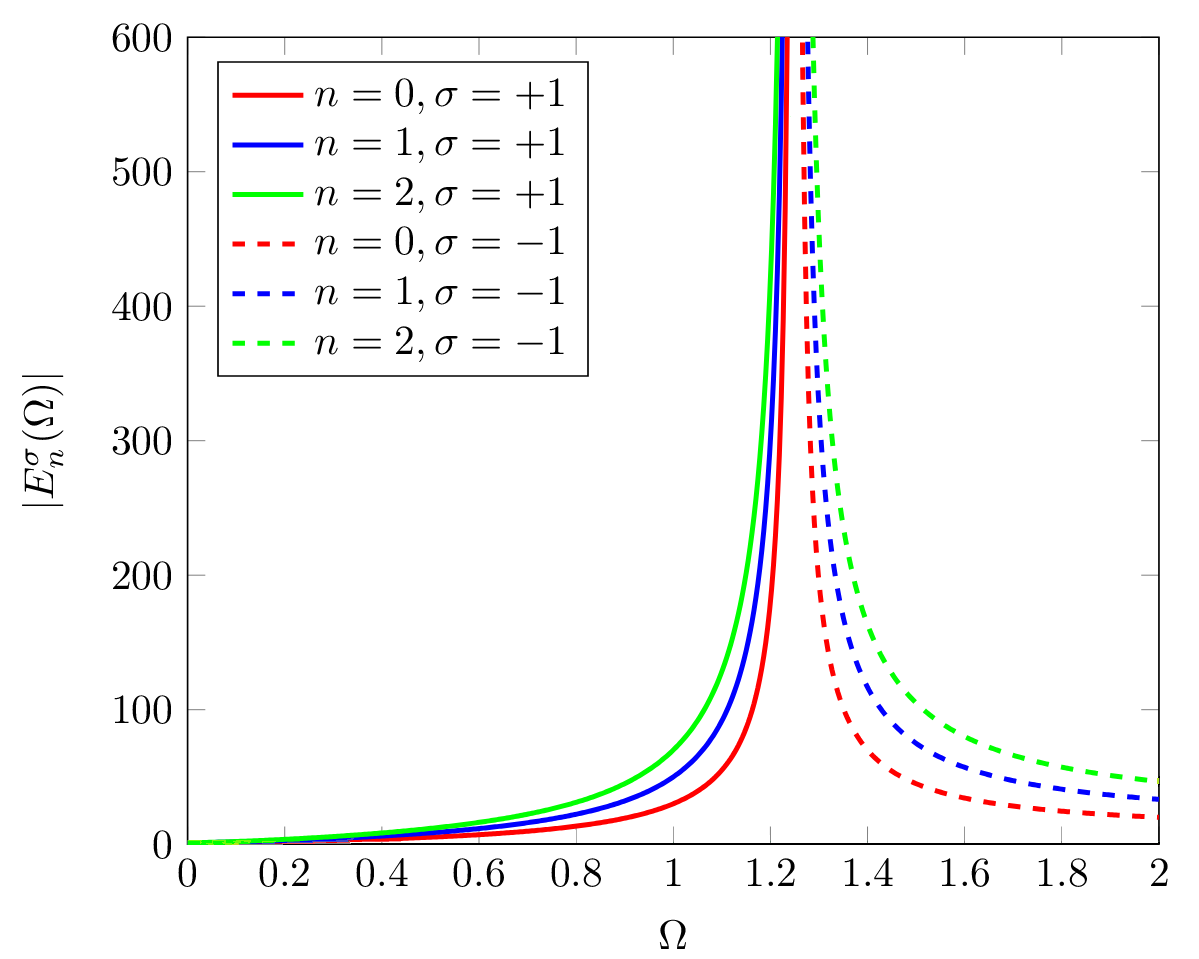}
\caption{Behavior of $\vert E^{\sigma}_n(\Omega)\vert$ versus $\Omega$ for three different values of $n$.}
\label{fig1}
\end{figure}

Now, let us graphically analyze the behavior of the spectrum (energy levels) as a function of the parameters $\Omega$, $\alpha$, and $\beta$ for three different values of $n$ (with $m_j=1/2$, i.e. we are considering the spectrum of setting 1). For simplicity, here we adopt $m_0=1$ (``unit mass''). In that way, in Fig. \ref{fig1} we have the behavior of the energies of the particle ($\sigma=+1$) and antiparticle ($\sigma=-1$) as a function of $\Omega$ for the ground state ($n = 0$) and the first two excited states ($n=1,2$), where we use $\beta=0.1$ and $\alpha=1/2$. So, according to this Figure, we see that the energies of the particle and antiparticle increase with the increase of $n$ (as it should be, otherwise, something would be wrong), that is, the energy difference between two consecutive levels is positive ($\Delta E_n=E_{n+1}-E_n>0$). Furthermore, the variation of the energies of the particle and antiparticle as a function of $\Omega$ presents an opposite behavior, for example, the energies of the particle increase with the increase of $\Omega$, tending to infinity at $\Omega=1.25$ (point of divergence), while the energies of the antiparticle increase with the decrease of $\Omega$, also tending to infinity at $\Omega=1.25$ (i.e. the energies only increase when $\Omega$ tends to the value of $1.25$). In other words, in the limit $\Omega\to 1.25$ (i.e. $\frac{4\Omega\beta}{\alpha}\to 1$), all energy levels of the particle as well as those of the antiparticle tend to infinity and the energy difference between them is practically zero ($\Delta E_n=0$). In this way, the energies of the particle vary (or are only allowed) in the range $0\leq\Omega\leq 1.25$ while those of the antiparticle vary (or are only allowed) in the range $1.25\leq\Omega<\infty$, respectively. 

\begin{figure}[!h]
\centering
\includegraphics[width=10.0cm]{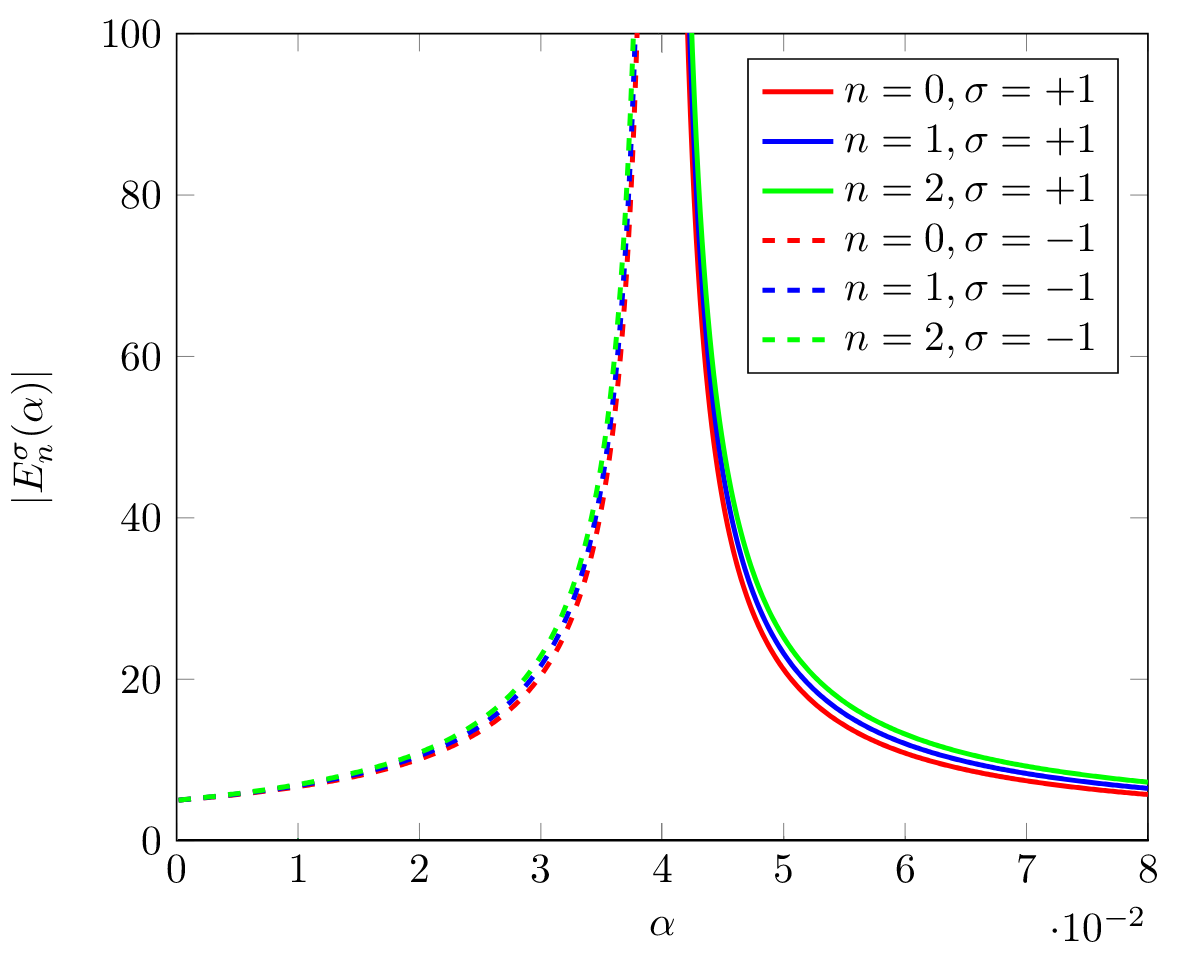}
\caption{Behavior of $\vert E^{\sigma}_n(\alpha)\vert$ versus $\alpha$ for three different values of $n$.}
\label{fig2}
\end{figure}

In Fig. \ref{fig2}, we have the behavior of the energies of the particle and antiparticle as a function of $\alpha$ for the ground state ($n = 0$) and the first two excited states ($n=1,2$), where we use $\Omega=\beta=0.1$. So, according to this Figure, we see that although the energies of the particle and antiparticle increase with the increase of $n$ (as it should be), the energy difference between two consecutive levels is practically zero for the antiparticle ($\Delta E_n\approx 0$), and slightly different from zero for the particle ($0<\Delta E_n\ll 1$). Furthermore, the variation of the energies of the particle and antiparticle as a function of $\alpha$ presents an opposite behavior, for example, the energies of the particle increase with the decrease of $\alpha$ (i.e. with the increase of conical curvature), tending to infinity at $\alpha=0.04$ (point of divergence), while the energies of the antiparticle increase with the increase of $\alpha$ (i.e. with the decrease of conical curvature), also tending to infinity at $\alpha=0.04$. Therefore, the energies only increase when $\alpha$ tends to the value of $0.04$. In other words, in the limit $\alpha\to 0.04$ (i.e. $\frac{4\Omega\beta}{\alpha}\to 1$), all energy levels of the particle as well as those of the antiparticle tend to infinity and the energy difference between them is practically zero ($\Delta E_n=0$). In this way, the energies of the particle vary (or are only allowed) in the range $0.04\leq\alpha<1$ while those of the antiparticle vary (or are only allowed) in the range $0<\alpha\leq 0.04$, respectively.

\begin{figure}[!h]
\centering
\includegraphics[width=10.0cm]{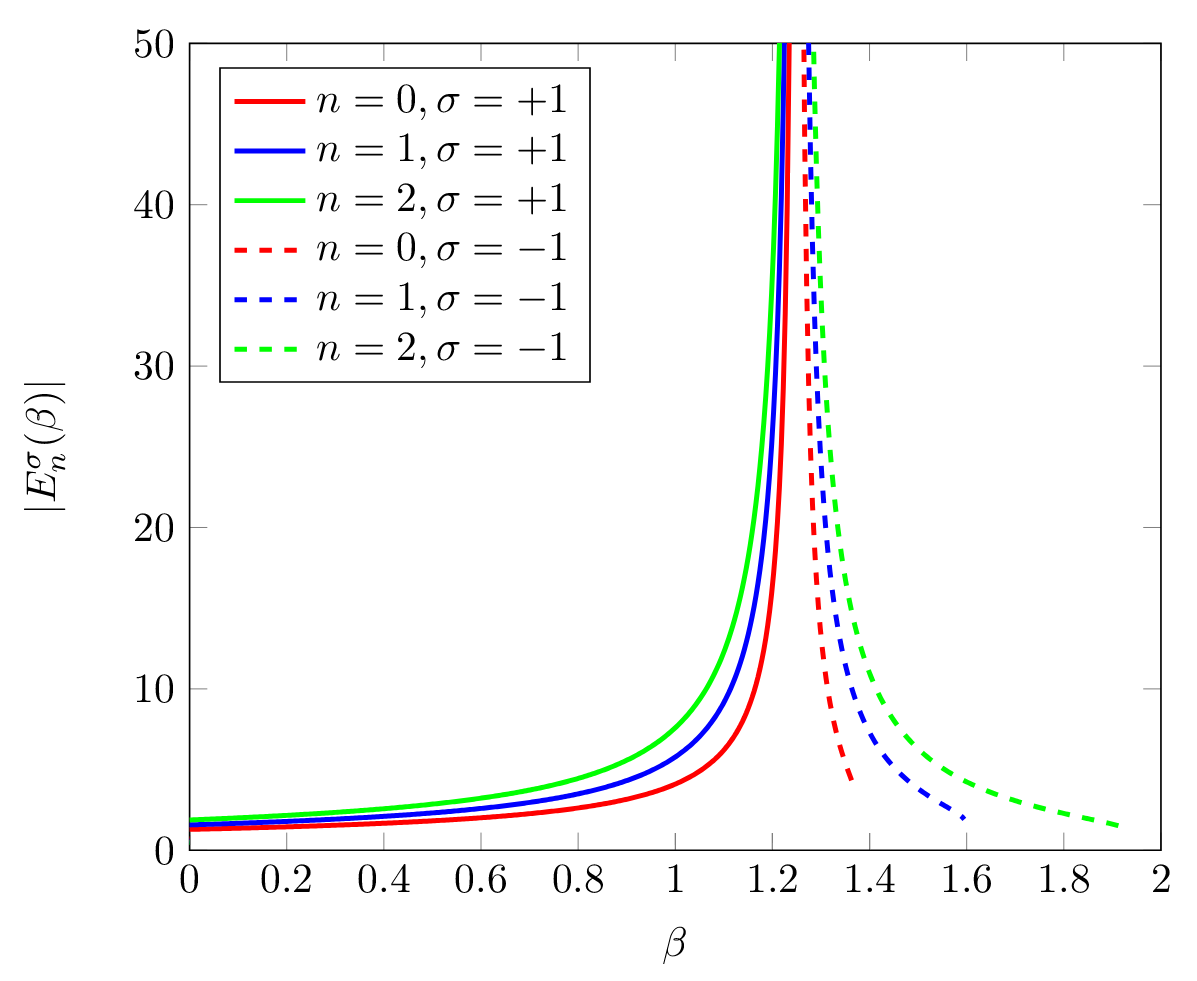}
\caption{Behavior of $\vert E^{\sigma}_n(\beta)\vert$ versus $\beta$ for three different values of $n$.}
\label{fig3}
\end{figure}

 Already in Fig. \ref{fig3}, we have the behavior of the energies of the particle and antiparticle as a function of $\beta$ for the ground state ($n = 0$) and the first two excited states ($n=1,2$), where we use $\Omega=0.1$ and $\alpha=1/2$. So, according to this Figure, we see that the energies of the particle and antiparticle present a behavior somewhat similar to that in Fig. \ref{fig1}, that is: increase with the increase of $n$ (as it should be), the energies of the particle increase with the increase of $\beta$ (tending to infinity at $\beta=1.25$), the energies of the antiparticle increase with the decrease of $\beta$ (tending to infinity at $\beta=1.25$), and the energies of the particle vary in the range $0\leq\beta\leq 1.25$ while those of the antiparticle vary in the range $1.25\leq\beta<2$, respectively.

\section{The nonrelativistic energy levels}\label{sec4}

Now, let us analyze the nonrelativistic limit (low-energy limit) of our results. To achieve this goal, we use a standard prescription (but with a modification) often used in literature to obtain the nonrelativistic limit of relativistic wave equations for massive particles \cite{Greiner}. So, in such prescription, we consider that most of the total energy of the system stays concentrated in the rest energy of the particle, as such $E=E^+\approx\frac{(m_0+\varepsilon)}{\left[1-\frac{4\Omega\beta}{\alpha}\left(\frac{\vert L_u \vert+L_u}{2L_u}\right)\right]}$ (a modified standard prescription), where $m_0 \gg \varepsilon$, $m_0 \gg 2\Omega N_u$, and $m_0 \gg \frac{\Omega}{2}$, respectively \cite{Greiner,GRG2024}. Indeed, for $\Omega=0$, $\beta=0$, $\frac{4\Omega\beta}{\alpha}\ll 1$, or $L_u<0$, we recover the original standard prescription (i.e. $E\approx m_0+\varepsilon$). Therefore, using the modified standard prescription in \eqref{spectrum}, we obtain the following nonrelativistic energy spectrum (nonrelativistic energy levels) for Dirac fermions (actually Pauli particles) in a spinning conical G\"odel-type spacetime
\begin{equation}\label{spectrum2}
\varepsilon_{n,m_j,s,u}=-\frac{\Omega}{2}+\frac{2\Omega N_u}{\sqrt{1-\frac{4\Omega\beta}{\alpha}\left(\frac{\vert L_u \vert+L_u}{2L_u}\right)}}, \ \ N_u=\left(n+\frac{1+su}{2}+\frac{\vert L_u\vert+L_u}{2\alpha}\right), \ \ L_u=m_j-\frac{su\alpha}{2}.
\end{equation}

So, we note that the nonrelativistic spectrum \eqref{spectrum2}, has some similarities and differences with the relativistic case (or relativistic spectrum of the fermion). For example, similar to the relativistic case, the spectrum \eqref{spectrum2}
\begin{itemize}
\item only admits positive energy states ($\varepsilon>0$), whose spectrum is also for a particle with spin-1/2: ``spin up'' ($s=+1$) or spin ``spin down'' ($s=-1$);
\item has its degeneracy broken due to the parameter $\alpha$;
\item depends on $n$, $m_j$, $\alpha$ and $\beta$ for $L_u$ or $m_j>0$ (however, does not depend on $s$ and $u$);
\item depends on $n$, $s$ and $u$ for $L_u<0$ or $m_j<0$ (however, does not depend on $m_j$, $\alpha$ and $\beta$);
\item increases infinitely when $\frac{4\Omega\beta}{\alpha}\to 1$ (for $L_u>0$).
\end{itemize}

However, unlike the relativistic case, the spectrum \eqref{spectrum2}
\begin{itemize}
\item does not depend on the rest mass $m_0$ (i.e. a highly massive particle and a less massive would have the same spectrum). In this way, the spectrum of the nonrelativistic particle is zero for $\Omega=0$;
\item depends linearly on the effective quantum number $N_u$ (or quantum numbers $n$ and $m_j$).
\end{itemize}

\begin{figure}[!h]
\centering
\includegraphics[width=10.0cm]{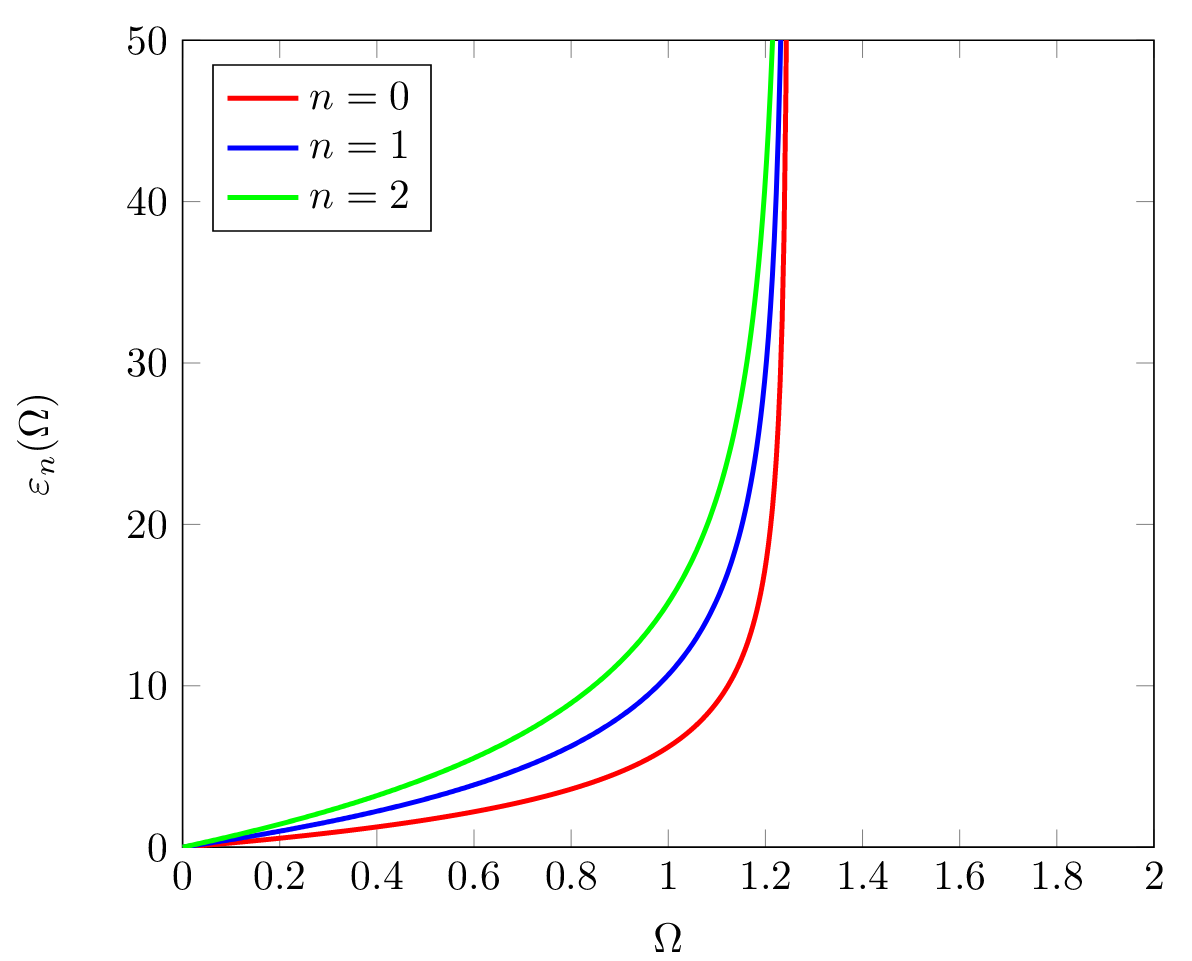}
\caption{Behavior of $\varepsilon_n(\Omega)$ versus $\Omega$ for three different values of $n$.}
\label{fig4}
\end{figure}

\begin{figure}[!h]
\centering
\includegraphics[width=10.0cm]{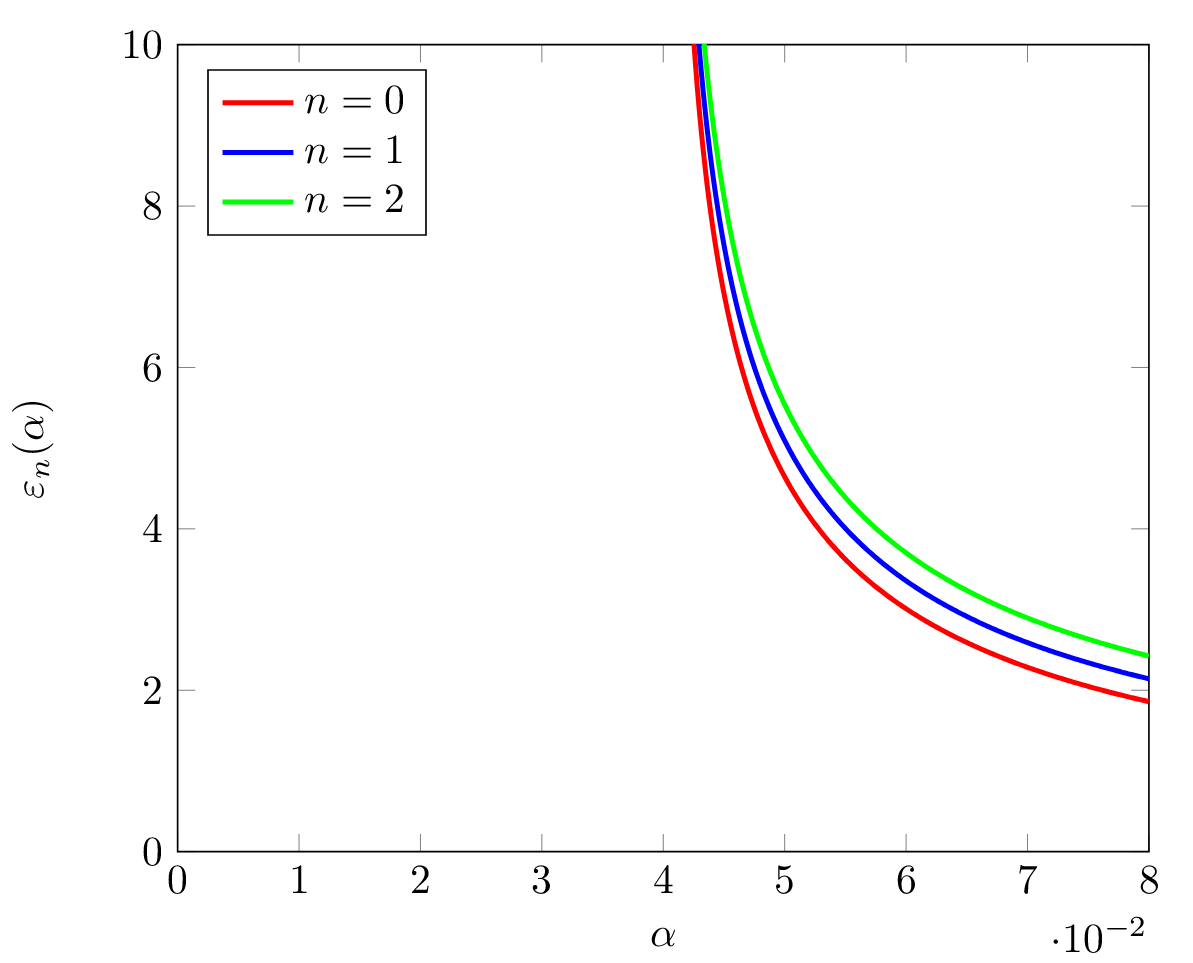}
\caption{Behavior of $\varepsilon_n(\alpha)$ versus $\alpha$ for three different values of $n$.}
\label{fig5}
\end{figure}

\begin{figure}[!h]
\centering
\includegraphics[width=10.0cm]{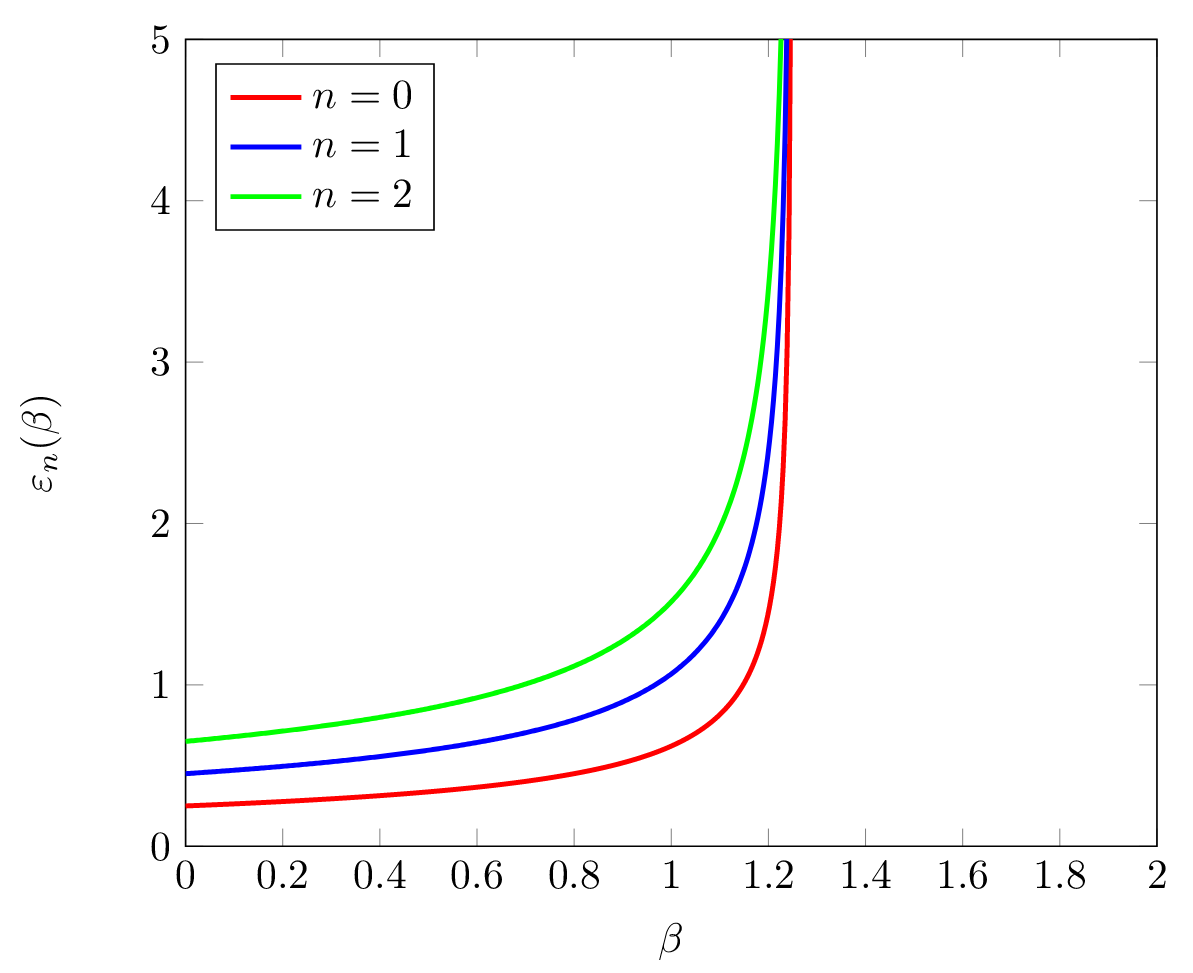}
\caption{Behavior of $\varepsilon_n(\beta)$ versus $\beta$ for three different values of $n$.}
\label{fig6}
\end{figure}

Now, let us graphically analyze the behavior of the spectrum as a function of $\Omega$, $\alpha$, and $\beta$ for three different values of $n$ (with $m_j=1/2$, i.e. we are considering the spectrum \eqref{spectrum2} for $L_u>0$). In particular, we will see that such behavior (i.e. the three graphs) is very similar to the relativistic case (the difference between the two cases is basically in the spectrum values). Therefore, in Fig. \ref{fig4} we have the behavior of the energies as a function of $\Omega$ for the ground state ($n = 0$) and the first two excited states ($n = 1, 2$), where we use $\beta= 0.1$ and $\alpha= 1/2$. So, according to this Figure, we see that the energies increase with the increase of $n$ (as it should be), and with the increase of $\Omega$ (tending to infinity at $\Omega= 1.25$). In Fig. \ref{fig5}, we have the behavior of the energies as a function of $\alpha$ for the ground state ($n = 0$) and the first two excited states ($n=1,2$), where we use $\Omega=\beta=0.1$. So, according to this Figure, we see that the energies increase with the increase of $n$ (as it should be), and with the decrease of $\alpha$ (i.e. with the increase of curvature), tending to infinity at $\alpha=0.04$. Already in Fig. \ref{fig6}, we have the behavior of the energies as a function of $\beta$ for the ground state ($n = 0$) and the first two excited states ($n = 1, 2$), where we use $\Omega= 0.1$ and $\alpha = 1/2$. So, according to this Figure, we see that the energies increase with the increase of $n$ (as it should be), and with the increase of $\beta$ (tending to infinity at $\beta= 1.25$).

\section{Conclusions}\label{sec5}

In this paper, we determine the relativistic and
nonrelativistic energy levels (or high and low energy spectrum) for Dirac fermions in a spinning conical G\"odel-type spacetime in $(2+1)$-dimensions, where such spacetime is the combination of the flat G\"odel-type spacetime with a spinning cosmic string and is modeled by three parameters: the vorticity parameter $\Omega$ of the G\"odel-type spacetime, and the curvature and rotation parameters $\alpha$ and $\beta$ of the spinning cosmic string. So, to achieve this goal, we work with the curved DE in polar coordinates $(t, r, \phi)$, where the formalism used to obtain the exact solutions of this equation was the tetrads formalism of GR. In addition, here we also consider the ``spin'' of the (2D) planar fermion, described by a parameter $s$, called the spin parameter ($s=+1$ is for the ``spin up'', and $s=-1$ is for the ``spin down''). In particular, this parameter was introduced into the literature with the goal of studying planar Dirac fermions interacting with external electromagnetic fields.

So, after solving (via variable change and asymptotic behavior) a second-order differential equation generated from two coupled first-order differential equations by the two components of the Dirac spinor, we obtain as results a generalized Laguerre equation and a quadratic polynomial equation for the total relativistic energy. Solving this polynomial equation, we obtain the relativistic energy spectrum for the fermion (particle) and antifermion (antiparticle), where we verified that such a spectrum explicitly depends on the quantum numbers $n$ and $m_j$ (the spectrum is quantized in terms of these quantum numbers), spin parameter $s$, spinorial parameter $u$, curvature parameter $\alpha$, rotation parameter $\beta$, and on the vorticity parameter $\Omega$, respectively. Additionally, a noteworthy observation is that the particle and antiparticle do not possess equal energies (i.e. the spectrum is not symmetrical). Therefore, this asymmetry in energy levels does not emphasize the equilibrium or equality between particle and antiparticle. In particular, the quantization of the spectrum is a direct result of the existence of $\Omega$, where such quantity acts as a kind of ``external field or potential'' (in fact, for $\Omega=0$, we have the spectrum of a truly free fermion/antifermion, given by $\pm m_0$). Furthermore, we also analyze the relativistic spectrum depending on the values of $m_j$. For example, we see that for $m_j>0$ (fermions/antifermion with positive angular momentum) the spectrum does not depend on $s$ and $u$ (the spectrum is the same regardless of the spin or spinor component chosen); however, it depends on both $n$, $m_j$, $\alpha$ and $\beta$. In this case, $\alpha$ breaks the degeneracy of the spectrum (i.e. the spectrum no longer has a well-defined degeneracy), and can increase infinitely in the limit $\frac{4\Omega\beta}{\alpha}\to 1$. In other words, depending on the values of $\Omega$, $\beta$ and $\alpha$, the energies can diverge. Already for $m_j<0$ (fermions/antifermion with negative angular momentum), we see that the spectrum depends on $s$ and $u$, as well as $n$ (i.e. now the spectrum depends on the spin and spinor component chosen); however, it no longer depends on both $m_j$, $\alpha$ and $\beta$. In this case, it is as if the fermion/antifermion ``lives only in a flat G\"odel-type spacetime'' (i.e. without the presence of a spinning cosmic string).

On the other hand, we also graphically analyzed the behavior of the spectrum $\vert E^\pm_n \vert$ (spectrum of the particle and antiparticle) as a function of $\Omega$, $\alpha$, and $\beta$ for three different values of $n$ ($n=0,1,2$ with $m_j= 1/2$). In that way, in graph $\vert E^\pm_n \vert$ versus $\Omega$ we see that the energies of the particle and antiparticle increase with the increase of $n$ (as it should be), however, the energies of the particle increase with the increase of $\Omega$ (tending to infinity at $\Omega=1.25$), while the energies of the antiparticle increase with the decrease of $\Omega$ (also tending to infinity at $\Omega=1.25$). In graph $\vert E^\pm_n \vert$ versus $\alpha$, we see that the energies of the particle and antiparticle increase with the increase of $n$ (as it should be), however, the energies of the particle increase with the decrease of $\alpha$ (tending to infinity at $\alpha=0.04$), while the energies of the antiparticle increase with the increase of $\alpha$ (also tending to infinity at $\alpha=0.04$). Already in graph $\vert E^\pm_n \vert$ versus $\beta$, we see that the energies of the particle and antiparticle present a behavior somewhat similar to the first graph, that is: increase with the increase of $n$ (as it should be), the energies of the particle increase with the increase of $\beta$ (tending to infinity at $\beta= 1.25$), and the energies of the antiparticle increase with the decrease of $\beta$ (tending to infinity at $\beta= 1.25$).

Finally, we also study the nonrelativistic limit (low-energy limit) of our results. Considering that most of the total energy of the system
stays concentrated in the rest energy of the particle, we obtain the nonrelativistic energy spectrum for Pauli particles in a spinning conical G\"odel-type spacetime. In particular, we note that this nonrelativistic spectrum has some similarities and differences with the relativistic case. For example, similar to the relativistic case, the nonrelativistic spectrum: only admits positive energy states (whose spectrum is also for a particle with spin-1/2), has its degeneracy broken due to the parameter $\alpha$, depends on $n$, $m_j$, $\alpha$ and $\beta$ for $m_j>0$ (but does not depend on $s$ and $u$), depends on $n$, $s$ and $u$ for $m_j<0$ (but does not depend on $m_j$, $\alpha$ and $\beta$), and increases infinitely when $\frac{4\Omega\beta}{\alpha}\to 1$. However, unlike the relativistic case, the spectrum does not depend on the rest mass $m_0$ (i.e. the mass of the particle has no influence on its spectrum), and depends linearly on the quantum numbers $n$ and $m_j$. Besides, we also graphically analyze the behavior of the nonrelativistic spectrum as a function of $\Omega$, $\alpha$, and $\beta$ for $n=0,1,2$ and $m_j=1/2$, where we note that such behavior is very similar to the relativistic case (the difference between the two cases is basically in the spectrum values).

So, as some future perspectives (topics for subsequent papers), we would like to offer a few suggestions, for example: 
\begin{itemize}
\item a) implementing the mathematical proof of the stability of the solutions (possibly using techniques from spectral theory or functional analysis);
\item b) exploring if there are any generalizations of the current model that could be analytically solved or perturbatively approximated, providing a broader context to its applicability (from cosmology to high energy physics);
\item c) supplementing analytical findings with numerical simulations to explore nonlinear dynamics or parameter regions where analytical solutions are challenging, to add a layer of validation, and also verify if new phenomena could emerge;
\item d) studying the problem of this paper with electromagnetic interactions (electromagnetic minimal and nonminimal couplings) via Hadamard’s method of descent (which is an alternative approach to the dimensional reduction of the DE).
\end{itemize}

\section*{Acknowledgments}

\hspace{0.5cm}
The author would like to thank the anonymous referees for the careful reading of the paper as well as for the remarkable suggestions and constructive comments that substantially helped in improving the quality of the paper. In addition, the author would also like to thank the Conselho Nacional de Desenvolvimento Cient\'{\i}fico e Tecnol\'{o}gico (CNPq) for financial support.

\section*{Data availability statement}

\hspace{0.5cm} This manuscript has no associated data or the data will not be deposited. [Author’s comment: There is no data because this paper is a theoretical study based on calculations of the relativistic and nonrelativistic energy levels for Dirac fermions in a spinning conical G\"odel-type spacetime.]


\begin{thebibliography}{99}
\section*{References}

\bibitem{Dirac1} P. A. M. Dirac, Proc. R. Soc. London A {\bf 117}, 610–624 (1928).

\bibitem{Dirac2} P. A. M. Dirac, Proc. R. Soc. London A {\bf 118}, 351–361 (1928).

\bibitem{Greiner} W. Greiner, {\it Relativistic quantum mechanics}, vol. 2 (Springer, Berlin, 2000). 

\bibitem{Bjorken} J. D. Bjorken, and S. D. Drell, {\it Relativistic quantum mechanics}, (Mcgraw-Hill College, New York, San Francisco, Toronto, London, 1964).

\bibitem{Grandy} W. T. Grandy, {\it Relativistic quantum mechanics of leptons and fields}, v. 41 (Springer Science and Business Media, 2012). 

\bibitem{Lesgourgues} J. Lesgourgues, and Pastor, S. Pastor, Phys. Rep. {\bf 429}, 307-379 (2006).

\bibitem{Studenikin} A. I. Studenikin, and I. V. Tokarev, Nucl. Phys. B {\bf 884}, 396-407 (2014).

\bibitem{Martin} B. R. Martin, and G. Shaw, {\it Particle physics}, (John Wiley and Sons, 2016)

\bibitem{griffiths} D. Griffiths, {\it Introduction to elementary particles}, (John Wiley and Sons, 2020).

\bibitem{Moshinsky} M. Moshinsky, and A. Szczepaniak, J. Phys. A: Math. Gen. {\bf 22}, L817 (1989).

\bibitem{Franco} J. A. Franco-Villafa{\~n}e et al, Phys. Rev. Lett. {\bf 111}, 170405 (2013).

\bibitem{Aharonov} Y. Aharonov, and A. Casher, Phys. Rev. Lett. {\bf 53}, 319 (1984).

\bibitem{Hagen1} C. Hagen, Phys. Rev. Lett. {\bf 64}, 503 (1990).

\bibitem{Hagen2} C. Hagen, Phys. Rev. Lett. {\bf 
64}, 2347 (1990).

\bibitem{Oliveira1} R. R. S. Oliveira, V. F. S. Borges, and M. F. Sousa, Braz. J. Phys. {\bf 49}, 801-807 (2019).

\bibitem{Oliveira2} R. R. S. Oliveira, R. V. Maluf, and C. A. S. Almeida, Ann. Phys. (N.Y.) {\bf 400}, 1-8 (2019).

\bibitem{Oliveira3} R. R. S. Oliveira et al, J. Phys. A: Math. Theor. {\bf 53}, 045304 (2020).

\bibitem{Oliveira4} R. R. S. Oliveira et al, Eur. Phys. J. Plus {\bf 134}, 495 (2019).

\bibitem{Boada} O. Boada et al, New J. Phys. {\bf 13}, 035002 (2011). 

\bibitem{Zhang} D. W. Zhang, Z. D. Wang, and S. L. Zhu, Front. Phys. {\bf 7}, 31-53 (2012).

\bibitem{Lamata} L. Lamata et al, Phys. Rev. Lett. {\bf 98}, 253005 (2007).

\bibitem{Gerritsma1} R. Gerritsma et al, Nature {\bf 463}, 68-71 (2010).

\bibitem{Schakel} A. M. Schakel, Phys. Rev. D {\bf 43}, 1428 (1991).

\bibitem{Fillion} F. Fillion-Gourdeau, S. MacLean, and R. Laflamme, Phys. Rev.A {\bf 95}, 042343 (2017).

\bibitem{Huerta} C. Huerta Alderete et al, Nat. Commun. {\bf 11}, 3720 (2020)

\bibitem{Bermudez1} A. Bermudez, M. A. Martin-Delgado, and A. Luis, Phys. Rev. A {\bf 77}, 063815 (2008).

\bibitem{Bermudez2} A. Bermudez, M. A. Martin-Delgado, and E. Solano, Phys. Rev. Lett. {\bf 99}, 123602 (2007).

\bibitem{Deriglazov} A. A. Deriglazov, Phys. Lett. A {\bf 376}, 309-313 (2012).

\bibitem{Casanova} J. Casanova et al, Phys. Rev. A {\bf 82}, 020101 (2010).

\bibitem{Gerritsma2} R. Gerritsma et al, Phys. Rev. Lett. {\bf 106}, 060503 (2011).

\bibitem{Beveren} E. Van Beveren, C.Dullemond, and T. A. Rijken, Phys. Rev. D {\bf 30}, 1103 (1984).

\bibitem{Becirevic} D. Becirevic, and A. Le Yaouanc, J. High Energy Phys. {\bf 1999}, 021 (1999).

\bibitem{Novoselov} K. S. Novoselov et al, Nature {\bf 438}, 197-200 (2005).

\bibitem{Gonzalez} J. Gonzalez, F. Guinea, and M. A. Vozmediano, Nucl. Phys. B {\bf 406}, 771-794 (1993).

\bibitem{McCann} E. McCann, and V. I. Fal’ko, J. Phys. Condens. Matter {\bf 16}, 2371 (2004).

\bibitem{Ahrens} S. Ahrens et al, New J. Phys. {\bf 17}, 113021 (2015).

\bibitem{Armitage} N. P. Armitage, E. J. Mele, and A. Vishwanath, Rev. Mod. Phys. {\bf 90}, 015001 (2018).

\bibitem{Dong} S. H. Dong, J. Phys. A: Math. Gen. {\bf 36}, 4977 (2003).

\bibitem{Setare} M. R. Setare, and S. Haidari, Phys. Scr. {\bf 81}, 065201 (2010).

\bibitem{Zou} X. Zou, L. Z. Yi, and C. S. Jia, Phys. Lett. A {\bf 346}, 54-64 (2005).

\bibitem{Adame} F. Domínguez-Adame, and M. A. González,  Europhys. Lett. (EPL) {\bf 13}, 193 (1990).

\bibitem{Guo} J. Y. Guo, and Z. Q. Sheng, Phys. Lett. A {\bf 338}, 90-96 (2005).

\bibitem{Jia} C. S. Jia et al, Eur. Phys. J. A {\bf 34}, 41-48 (2007).

\bibitem{SousaOliveira} R. R. Sousa Oliveira, G. Alencar, and R. R. Landim, Phys. Scr. {\bf 99}, 035226 (2024).

\bibitem{Oliveira5} R. R. S. Oliveira, and R. R. Landim, Thermodynamic properties of the noncommutative Dirac oscillator with a permanent electric dipole moment, arXiv preprint: 2212.10339 (2022).

\bibitem{Oliveira6} R. R. Sousa Oliveira, and R. R. Landim, Phys. Scr. {\bf 99}, 045917 (2024).

\bibitem{Guvendi} A. Guvendi, Eur. Phys. J. C {\bf 84}, 185 (2024).

\bibitem{Ning} W. Ning et al, Npj Quantum Inf. {\bf 9}, 99 (2023).

\bibitem{Jakubassa} D. H. Jakubassa-Amundsen, J. Phys. G: Nucl. Part. Phys. {\bf 51}, 035105 (2024).

\bibitem{b1} K. Bakke, and H. Belich, Universe {\bf 9}, 462 (2023).

\bibitem{b2} K. Bakke, and H. Belich, Europhys. Lett. (EPL) {\bf 141}, 40004 (2023).

\bibitem{Ahmed} F. Ahmed, and A Guvendi, Nucl. Phys. B {\bf 1000}, 116470 (2024).

\bibitem{ARXIV} Relativistic and nonrelativistic Landau levels for Dirac fermions in the cosmic string spacetime in the context of rainbow gravity, arXiv preprint: 2403.01366 (2024).

\bibitem{Godel} K. G\"odel, Rev. Mod. Phys. {\bf 21}, 447 (1949).

\bibitem{Deszcz} R. Deszcz et al, Int. J. Geom. Methods Mod. Phys. {\bf 11}, 1450025 (2014).

\bibitem{Gleiser} R. J. Gleiser et al, Class. Quantum Grav. {\bf 23}, 2653 (2006).

\bibitem{Barrow} J. D. Barrow, and M. P. Dabrowski, Phys. Rev. D {\bf 58}, 103502 (1998).

\bibitem{Buser} M. Buser, E. Kajari, and W. P. Schleich, New J. Phys. {\bf 15}, 013063 (2013).

\bibitem{Kerner} R. Kerner, and R. B. Mann, Phys. Rev. D {\bf 75}, 084022 (2007).

\bibitem{Rebouças} M. J. Rebouças, and J. Tiomno, Phys. Rev. D {\bf 28}, 1251 (1983).

\bibitem{Figueiredo} B. D. B. Figueiredo, I. D. Soares, and J. Tiomno, Class. Quantum Gravity {\bf 9}, 1593 (1992).

\bibitem{Drukker} N. Drukker, B. Fiol, and J. Simón, J. Cosmol. Astropart. Phys. {\bf 2004}, 012 (2004).

\bibitem{Boyda} E. K. Boyda et al, Phys. Rev. D {\bf 67}, 106003 (2003).

\bibitem{Kibble} T. W. B. Kibble, J. Phys. A {\bf 19}, 1387 (1976).

\bibitem{Vilenkin} A. Vilenkin, Phys. Rep. {\bf 121}, 263 (1985).

\bibitem{Bezerra} V. B. Bezerra, J. Math. Phys. {\bf 38}, 2553-2564 (1997).

\bibitem{Mello} E. R. B. Mello, J. High Energy Phys. {\bf 2004}, 016 (2004)

\bibitem{Oli1} R. R. S. Oliveira, Gen. Relativ. Gravit. {\bf 51}, 120 (2019).

\bibitem{Oli2} R. R. S. Oliveira, Eur. Phys. J. C {\bf 79}, 725 (2019).

\bibitem{Oli3} R. R. S. Oliveira, Gen. Relativ. Gravit. {\bf 52}, 88 (2020).

\bibitem{Oli4} R. R. S. Oliveira, G. Alencar, and R. R. Landim, Gen. Relativ. Gravit. {\bf 55}, 15 (2023).

\bibitem{INDIANA} R. R. S. Oliveira, R. V. Maluf, and C. A. S. Almeida, Indian J. Phys. {\bf 98}, 1-9 (2024).

\bibitem{Cui} Y. Cui et al. J. High Energy Phys. {\bf 2019}, 1 (2019).

\bibitem{Auclair} P. Auclair et al, J. Cosmol. Astropart. Phys. {\bf 2020}, 034 (2020).

\bibitem{Blasi} S. Blasi, V. Brdar, and K. Schmitz, Phys. Rev. Lett. {\bf 126}, 041305
(2021).

\bibitem{Liu} Y. Liu and al, Nature {\bf 589}, 381–385 (2021).

\bibitem{Katanaev} M. Katanaev, and I. Volovich, Ann. Phys. (N.Y.) {\bf 216}, 1 (1992).

\bibitem{Ardeshana} B. Ardeshana, et al, AIMS Mater. Sci. {\bf 4}, 1010–1028 (2017).

\bibitem{Vozmediano} M. A. H. Vozmediano, M. I. Katsnelson, F. Guinea, Phys. Rep. {\bf 496}, 109 (2010).

\bibitem{Cortijo} A. Cortijo, M. A. H. Vozmediano, Europhys. Lett. {\bf 77}, 47002 (2007).

\bibitem{Rozhkov} M. A. Rozhkov et al, Low Temp. Phys. {\bf 44}, 918-924 (2018).

\bibitem{Carv} J. Carvalho, A. M. de Carvalho, and C. Furtado, Eur. Phys. J. C {\bf 74}, 2935 (2014)

\bibitem{GRG2024} R. R. S. Oliveira, Gen. Relativ. Gravit. {\bf 56}, 30 (2024).

\bibitem{Montigny} M. de Montigny, S. Zare, and H. Hassanabadi, Gen. Relativ. Gravit. {\bf 50}, 1-24 (2018).

\bibitem{Garcia} G. Q. Garcia, J. D. S. Oliveira, and C. Furtado, Int. J. Mod. Phys. D {\bf 27}, 1850027 (2018).

\bibitem{Vitória} R. L. L. Vitória, C. Furtado, and K. Bakke, Eur. Phys. J. C {\bf 78}, 1-5 (2018).

\bibitem{Eshghi} M. Eshghi, and M. Hamzavi, Eur. Phys. J. C {\bf 78}, 522 (2018).

\bibitem{Ahmed1} F. Ahmed, Eur. Phys. J. C {\bf 79}, 534 (2019).

\bibitem{Ahmed2} F. Ahmed, Eur. Phys. J. C {\bf 79}, 104 (2019).

\bibitem{Sedaghatnia} P. Sedaghatnia, H. Hassanabadi, and F. Ahmed, Eur. Phys. J. C {\bf 79}, 541 (2019).

\bibitem{Mustafa1} O. Mustafa, Phys. Scr. {\bf 98}, 015302 (2022).

\bibitem{Mustafa2} O. Mustafa, Eur. Phys. J. Plus {\bf 138}, 21 (2023).

\bibitem{Som} M. M. Som, and A. K. Raychaudhuri, Proc. R. Soc. A {\bf 304}, 81 (1968).

\bibitem{Horowitz} G. T. Horowitz, and A. A. Tseytlin, Phys. Rev. D {\bf 51}, 2896 (1995).

\bibitem{Russo1} J. G. Russo, and A. A. Tseytlin, Nucl. Phys. B {\bf 448}, 293-328 (1995).

\bibitem{Russo2} J. G. Russo, and A. A. Tseytlin, Nucl. Phys. B {\bf 454}, 164-184 (1995).

\bibitem{Harmark} T. Harmark, and T. Takayanagi, Nucl. Phys. B {\bf 662}, 3-39 (2003).

\bibitem{Cho} Y. M. Cho, D. H. Park, and C. G. Han, Phys. Rev. D {\bf 43}, 1421 (1991).

\bibitem{Angelone} G. Angelone et al, J. Phys. A: Math. Theor. {\bf 56}, 065201 (2023).

\bibitem{Lawrie} I. D. Lawrie, {\it A Unified Grand Tour of Theoretical Physics}, vol. 3
(CRC Press, London, 2012).

\end{thebibliography}
\end{document}